\documentclass[a4paper,11pt]{article}

\usepackage{contribution}

\newcommand{\weblink}[2][]{%
    \ifthenelse{\equal{#1}{}}%
    {\textnormal{\url{#2}}}%
    {\textnormal{\href{#2}{#1}}}%
}

\newcommand{\acknowledgements}[1]{%
  \bigskip\bigskip
  \textsf{\textbf{\Large Acknowledgements}} \\[2ex]
  {#1}
  \bigskip
}

\def\beq{\begin{equation}}
\def\eeq#1{\label{#1}\end{equation}}
\def\eeqn{\end{equation}}

\def\beqa{\begin{eqnarray}}
\def\eeqa#1{\label{#1}\end{eqnarray}}
\def\eeqan{\end{eqnarray}}

\let\bar=\overbar

\def\Dslash{\not{\hbox{\kern-4pt $D$}}}
\def\dslash{\not{\hbox{\kern-2pt $\del$}}}

\def\msb{{\bar{\ssstyle M \kern -1pt S}}}

\newcommand{\contribution}[7][]{%
  \clearpage
  \thispagestyle{plain}
  \ifthenelse{\equal{#1}{}}
  {\hypersetup{pdftitle={#2}}}
  {\hypersetup{pdftitle={#1}}}
  \hypersetup{pdfauthor={{#3} {#4}}}
  {\centering\normalfont\LARGE\bfseries\sffamily #2 \par\nobreak}
  \lhead{}
  \chead{%
    \textit{\footnotesize XIV International Conference on Hadron Spectroscopy
      (\weblink[\textit{hadron2011}]{http://www.hadron2011.de}), 13-17 June 2011, Munich, Germany}%
  }
  \rhead{}
  \bigskip
  \begin{center}
    {#3} {#4}\ifthenelse{\equal{#6}{}}{}{\footnote{\weblink[#6]{mailto:#6}}}
    \ifthenelse{\equal{#7}{}}{}{#7} \\
    \textit{#5}
  \end{center}
  \bigskip
}

\renewcommand{\abstract}[1]{%
  \begin{center}
    \begin{minipage}{0.85\textwidth}
      \begin{footnotesize}
        #1
      \end{footnotesize}
    \end{minipage}
  \end{center}
  \bigskip
}

\begin{document}

%
%
%
%
%
{  


%

\vspace{-0.5cm}
\contribution[Study of $dp$ and $dd$ Collisions]  
{Study of Deuteron-Proton and Deuteron-Deuteron Collisions at
Intermediate Energies}  
{Nadezhda}{Ladygina} 
{Laboratory of High Energy Physics\\
Joint Institute for Nuclear Researches \\
141980 Dubna, RUSSIA}  
{}  
{}

\vspace{-0.5cm}
During a few  decades hadronic  reactions with a participation of the light nuclei were extensively investigated
 at  the energies of  few hundred MeV. These processes are the simplest 
examples of the hadron nucleus collision that is why the interest to 
this reaction is justified. A number  of experiments 
 is aimed at getting some information
about the deuteron or helium wave functions as well as nucleon-nucleon amplitudes
from the scattering observables.

In this paper I consider two reactions. The first of them is the $dp$-elastic 
scattering in the deuteron energy range between 500 MeV and 2 GeV. The second 
reaction is the $dd\to {^3He~n}$ at the energies from 200 MeV up to 520 MeV.
At these energies the methods based on the solution of the Faddeev equations 
are unusable. I start from the AGS-equations \cite{ags},\cite{ags4} and iterate
their over nucleon-nucleon $t$-matrix.  The plane-wave-impulse-approximations (PWIA),
single scattering (SS) and double scattering (DS) contributions are taken into consideration
for the $dp\to dp$ process. For the $dd\to {^3He}~n$ reaction I consider the
one-nucleon-exchange (ONE) and single scattering terms.  The applied approach was presented
in details in refs.\cite{japh}, \cite{euro}.  
\begin{figure}[htb]
\begin{minipage}[t]{0.4\textwidth}
\hspace{-1cm}
\includegraphics[height=1.5in, width=3.in]{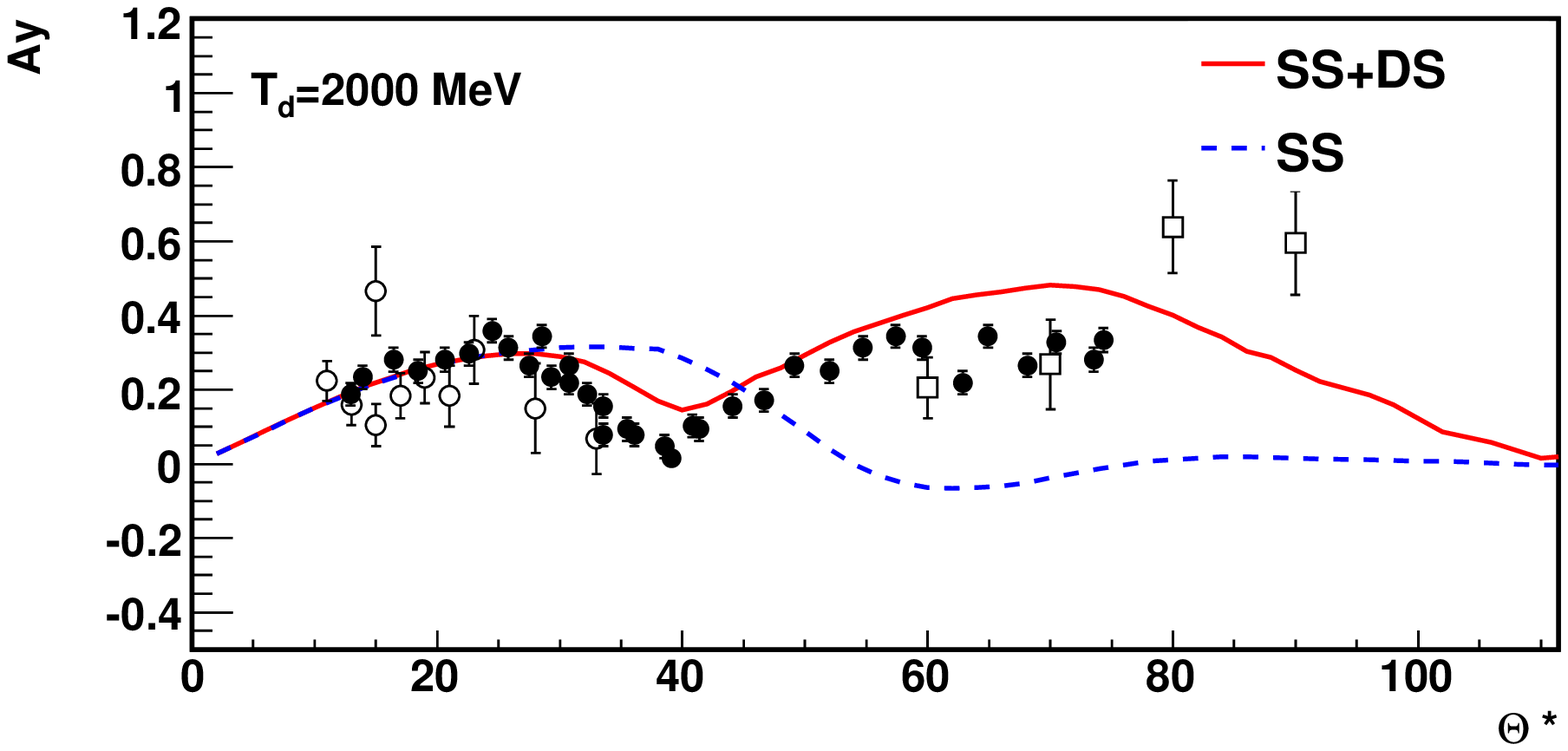}
\end{minipage}
\hfill
\begin{minipage}[t]{0.4\textwidth}
\vspace{-4cm}
\hspace{-1.5cm}
\includegraphics[height=1.5in, width=3.in]{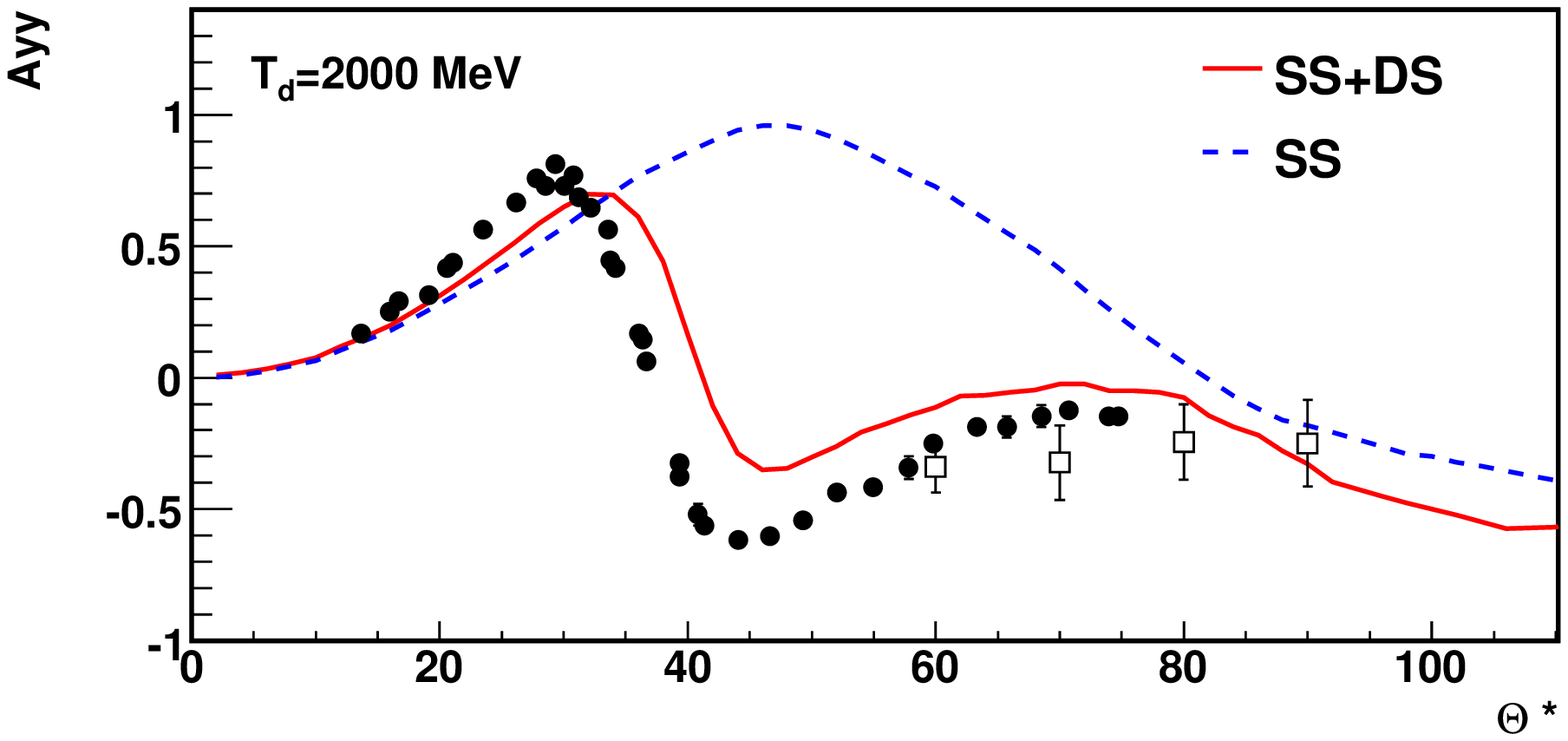}
\end{minipage}
\vspace{-0.5cm}
\caption{The vector,$A_y$, and tensor, $A_{yy}$, analyzing powers of the deuteron in the $dp\to dp$. 
 The data are taken from: $\bullet$ \protect\cite {sacle}; 
$\circ$  LHE JINR, hydrogen bubble chamber experiment, and
$\square$ LHE JINR Nuclotron, talk by P.Kurilkin given at this conference.}
\end{figure}

The results of the calculations for the $dp$- elastic scattering are given in Fig.1.
The angular dependencies of the vector and tensor analyzing powers are presented
at the deuteron 
energy of 2 GeV. Here, the solid curves correspond to the results of calculations
including both the single scattering and double scattering terms.
\begin{figure}[htb]
\begin{minipage}[t]{0.4\textwidth}
\hspace{-1cm}
\includegraphics[height=1.5in, width=3.in]{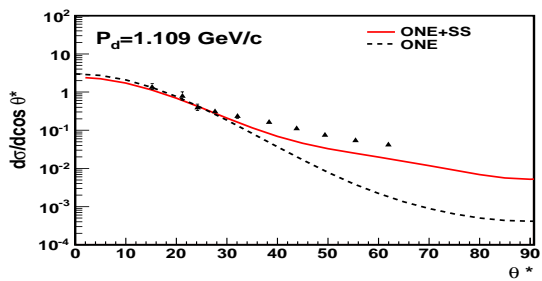}
\caption{The differential cross section of the $dd\to {^3He~n}$. 
The data are taken from \protect\cite {bizard}.
}
\end{minipage}
\end{figure}
\begin{figure}
\hfill
\begin{minipage}[t]{0.4\textwidth}
\vspace{-6.5cm}
\hspace{-1.5cm}
\includegraphics[height=1.5in, width=3.in]{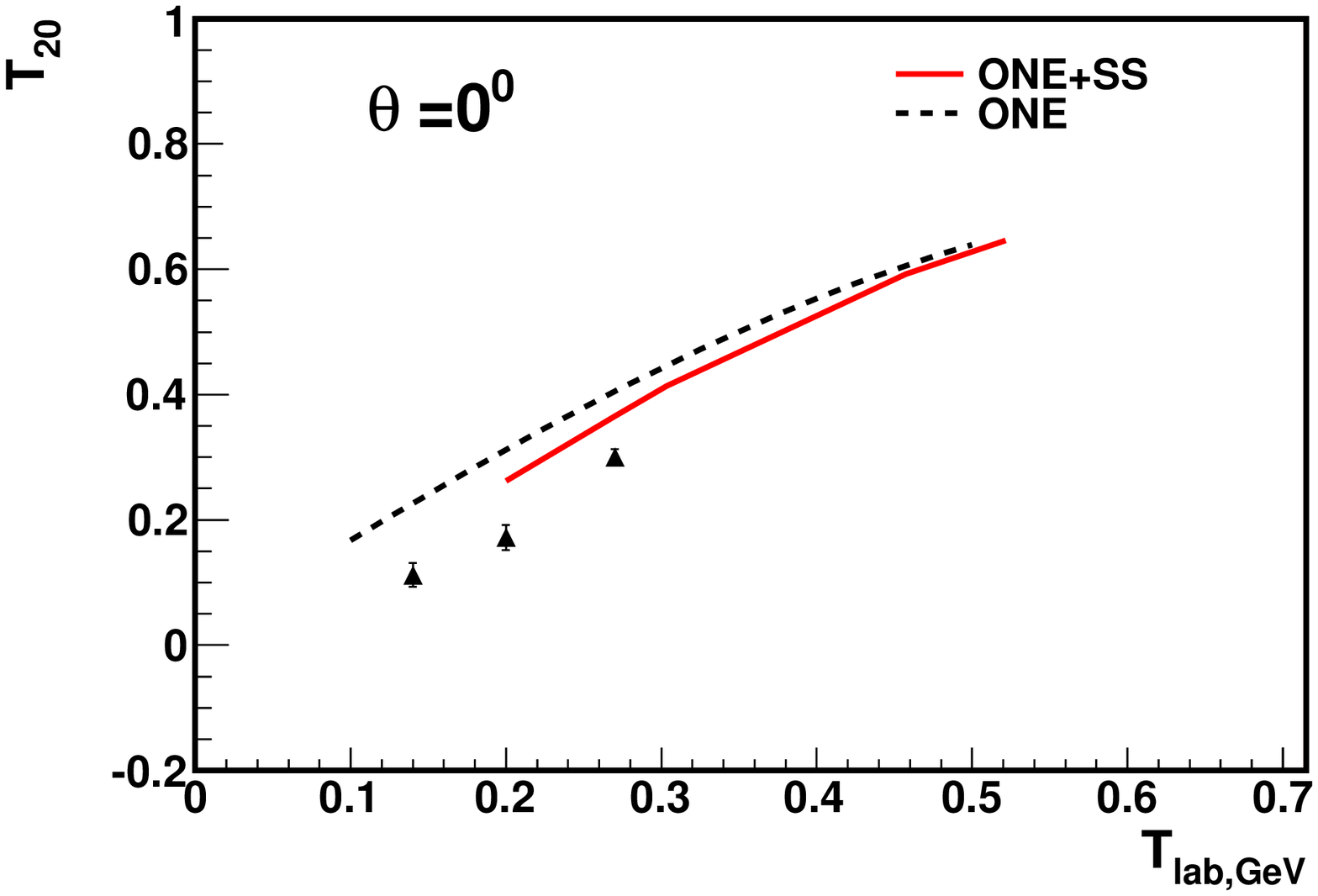}
\vspace{-0.5cm}
\hspace*{-1.5cm}
\caption{The tensor analyzing power in the $dd\to {^3He~n}$. 
The data are taken from  \protect\cite {T20}. }
\end{minipage}
\end{figure}
 The
 dashed curves
are the results taking of account only SS-contribution.
One can see the inclusion of the
DS-term  significantly improves the agreement between the theoretical predictions and experimental
data, especially, for the tensor analyzing power. 

The differential cross section of the $dd\to {^3He}~n$ reaction is presented in Fig.2 at laboratory
momentum of 1.109 GeV/c \cite{bizard}. Also the energy
dependence of the  $T_{20}$ has been obtained at zero scattering angle in the energy range between 200 MeV and
520 MeV (Fig.3).   
The results demonstrate the significant improvement an agreement of the data and
theory predictions, especially, for non forward angles, when the single scattering 
term is included.  It allows us to regard this approach as a next step
towards a solution of the four-nucleon problem. 

\vspace{-0.6cm}
\acknowledgements{%
This work has been supported by the Russian Foundation for Basic Research
under grant  $N^{\underline 0}$  10-02-00087a.
}

%

}  


\end{document}